\documentstyle[prb,aps,epsf,twocolumn]{revtex}
\newcommand{\eq}{\begin{equation}}
\newcommand{\ee}{\end{equation}}
\newcommand{\s}{{\sigma}}

\newcommand{\vq}{{\vec{q}}}

\newcommand{\vQ}{{\vec{Q}}}

\def\o{{\omega}}

\def\ad{a^{\dagger}}

\def\ua{\uparrow}
\def\da{\downarrow}
\def\eqa{\begin{eqnarray}}
\def\eea{\end{eqnarray}}
\def\prl{{Phys. Rev. Lett.}}
\def\prb{{Phys. Rev. {\bf B}}}

\def\jpsj{{Jour. Phys. Soc. Japan\ }}

\begin{document}
\draft
\flushbottom
\twocolumn[
\hsize\textwidth\columnwidth\hsize\csname @twocolumnfalse\endcsname
\title{Hall Crystal States at $\nu=2$ and Moderate Landau Level Mixing}
\author{  Ganpathy Murthy}
\address{
Department of Physics and Astronomy, University of Kentucky, Lexington, KY 40506}
\date{\today}
\maketitle
\tightenlines
\widetext
\advance\leftskip by 57pt
\advance\rightskip by 57pt

\begin{abstract}
The $\nu=2$ quantum Hall state at low Zeeman coupling is well-known to
be a translationally invariant singlet if Landau level mixing is
small. At zero Zeeman interaction,  as Landau
level mixing increases, the translationally
invariant state becomes unstable to an inhomogeneous state. This is
the first realistic example of a full Hall crystal, which shows the
coexistence of quantum Hall order and density wave order. The full Hall
crystal differs from the more familiar Wigner crystal by a topological
property, which results in it having only linearly dispersing
collective modes at small $q$, and no $q^{3/2}$ magnetophonon. I
present calculations of the topological number and the collective
modes.

\end{abstract}
\vskip 1cm

]
\narrowtext
\tightenlines
Integer quantum Hall states are some of the best understood, because
single-particle approximations such as Hartree-Fock (HF) work very
well for filled Landau levels (for an overview of the rich physics in
these systems, see ref. [1]). The three important parameters that
describe an integer state are the cyclotron frequency $\o_c$, the
Zeeman coupling $E_Z$, and the dimensionless parameter characterizing
the strength of the electron-electron interaction $r_s$. For $\nu=2$
we have $r_s={e^2\over \varepsilon l_0 \hbar\o_c}$.  As $r_s$ becomes
large  Landau level mixing increases.

We will work in the limit $E_Z=0$, which is an excellent approximation
for GaAs based systems, due to the reduction of the $g$ factor, and
the enhancement of the cyclotron frequency by band effects. In this
limit, $r_s$ survives as the only dimensionless parameter
characterizing the system. The Hamiltonian in standard notation is
\eq
H=\sum n\hbar\o_c \ad_{\s,n,X} a_{\s,n,X} + {1\over 2L^2}\sum v(q) :\rho(\vq)\rho(-\vq):
\ee
where $(\s,n,X)$ are spin, Landau level, and degeneracy indices for
single-particle states in the Landau gauge, $a,\ad$ are the fermion
operators which destroy and create electrons in these single-particle
states, and $\rho(\vq)$ is the density operator
\eq
\rho(\vq)=\sum\limits_{\s n_1n_2X} e^{-iq_x X} \rho_{n_1n_2}(\vq)\ad_{\s,n_1,X-q_yl_0^2/2}a_{\s,n_2,X+q_yl_0^2/2}
\ee
Here $l_0$ is the magnetic length, and $\rho_{nn'}(\vq)$ is a matrix
element.

What are the possible ground states of the system?  At $\nu=2$ the
simplest possiblities are (i) Fill the $n=0,
\ua$ and $n=0,
\da$ Landau levels to form the singlet state, or 
(ii) Fill $n=0, 1$ for the $\ua$ spins only to form the fully
polarized state. One can check that in the HF approximation, the fully
polarized state becomes lower in energy than the singlet state for
$r_s\ge2.12$ for the Coulomb interaction, a transition first pointed
out in a slightly different context by Giuliani and
Quinn\cite{giuliani}. Finite thickness effects are modelled by an
interaction 
$v(q)={e^2\over \varepsilon q} e^{-\lambda q}$ 
where the length $\lambda$ is related to the sample thickness.  This
modifies the Coulomb interaction at large $q$ and pushes the critical
$r_s$ for the singlet-fully polarized transition higher.

Translationally invariant states cannot take advantage of Landau level
mixing (in HF), while inhomogeneous states can. Inhomogeneous states
have been the subject of intense investigation in the early eighties
in HF\cite{yosh-fuku,fuku,yosh-lee,macd,wc} in the context of their
possible relevance to the fractional quantum Hall effect and the
high-field Wigner crystal. In HF, such states are described by nonzero
expectation values
\eq
\Delta_{\s n,\s' n'}(\vQ)={2\pi l_0^2\over L^2}\sum e^{-iQ_x X} \langle \ad_{\s n,X-Q_yl_0^2/2} a_{\s' n',X+Q_yl_0^2/2}\rangle
\ee
where $L^2$ is the area of the system, and $\vQ$ are the set of
reciprocal vectors of some lattice.  Since we have two spin flavors,
we can have states with spin mixing or not. Using these expectation
values, one decouples the interaction term. One then performs a
sequence of canonical transformations\cite{yosh-lee,macd} which
reduces the problem to diagonalizing a matrix for every point in the
magnetic Brillouin zone. The dimension of this matrix is connected to
the number of flux quanta per unit cell, and the number $n_{LL}$ of
LLs kept (typically I keep $n_{LL}=10$ levels). If the flux per unit
cell is $\phi=p\phi_0$ (here $\phi_0$ is the flux quantum), then each
Landau level breaks up into $p$ nonoverlapping subbands. I have
examined density wave states with two and three flux quanta per unit
cell, with varying number of majority-spin subbands occupied, and for
the square and triangular lattices. In the regime $3\le r_s\le 9$ that
I will mostly focus on, the triangular lattice with two flux quanta
per unit cell, no spin-mixing, and a total polarization of half the
maximal polarization turns out to have the lowest energy among all the
density wave states that I studied. Anticipating the result that this
is a Hall crystal, I will call this the partially polarized Hall
crystal (PPHC). This state is the integer analog of the partially
polarized density wave state for $\nu=2/5$ that I have
proposed\cite{murthy} to explain the direct spin polarization
measurements of Kukushkin {\it et al}\cite{kukush}.

Figure 1 presents the results for the ground state energy for the
translationally invariant singlet (S) state, the fully polarized (FP)
state, and for the triangular PPHC state for the pure Coulomb
interaction ($\lambda=0$) and for a sample with finite thickness
($\lambda=0.4l_0$). For $\lambda=0$ the singlet is the lowest state
among these for $r_s\le2.12$, while the PPHC state becomes the lowest
energy state for $r_s\ge 6$. For $\lambda=0.4l_0$ there is a direct
transition from the S state to the PPHC state at $r_s\approx 4.5$, and
the FP state is never the ground state.

\begin{figure}
\narrowtext
\epsfxsize=2.4in\epsfysize=2.4in
\hskip 0.3in\epsfbox{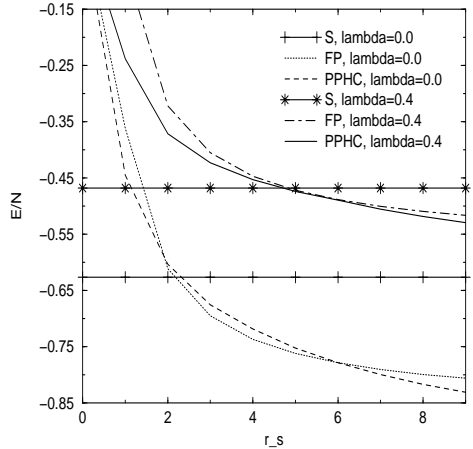}
\vskip 0.15in
\caption{Ground state energy per particle as a function of $r_s$ 
for the singlet, fully polarized, and PPHC states for $\lambda=0.0$ 
and $\lambda=0.4$. 
The solid line is the singlet energy for $\lambda=0.0$, while the 
dotted line is the singlet energy for $\lambda=0.4$. 
\label{fig1}}
\end{figure}

The PPHC state is distinguished from the S and the FP states by a
density wave and a partial polarization. Another very important
topological property of a crystal state was elucidated by Thouless and
co-workers\cite{thouless}.  Suppose one considers a set of
noninteracting electrons in the presence of a periodic potential. Then
one can ask how much charge is transported when the lattice potential
is adiabatically dragged by a lattice translation vector. It was
shown\cite{thouless} that in the thermodynamic limit, if the chemical
potential lies in a gap, this charge transported is quantized, and
characterized by an integer Chern index. In integer quantum Hall
systems with a periodic potential, two integers, the quantized Hall
conductance, and the above-mentioned Chern number, characterize each
state.

More generally, one can ask whether crystalline and quantized Hall
order can coexist, as was suggested by the cooperative ring-exchange
theory\cite{ring-ex}. This question was answered in the affirmative
and the physical significance of the second integer was clarified by
Tesanovic, Axel, and Halperin\cite{hall-crystal}. Given the two
integers, the average density obeys the equation
\eq
{\bar \rho}=n_H\rho_{\phi}+n_C A_0^{-1}
\ee
where $n_H=h\sigma_{yx}/e^2$ is the integer characterizing the Hall
conductance, $\rho_{\phi}$ is the density of flux quanta, $A_0$ is the
area of the unit cell, and $n_C$ is the Chern number describing the
adiabatic transport of charge. The usual quantum Hall states have
$n_H\ne0$, but $n_C=0$, while the usual Wigner crystals in the quantum
Hall regime have $n_H=0$, but $n_C\ne 0$.

Tesanovic {\it et al}\cite{hall-crystal} considered the case when
there was a density wave, but with $n_C=0$, which they called a {\it
full Hall crystal}. They also labelled states with nonzero values of
both integers as {\it partial Hall crystals} (not to be confused with
partial polarization!). They explicitly constructed a (rotationally
anisotropic) interaction with two-body and four-body parts for which
they were able to show that the ground state was a full Hall
crystal. One of their most important results concerns the low-energy
collective modes of the various states\cite{hall-crystal}. It has long
been known that the Wigner crystal has a single gapless magnetophonon
collective mode with a dispersion of $\o\propto q^{3/2}$ (for the
long-range Coulomb interaction). This arises because the magnetic
field mixes the usual ($B=0$) linearly dispersing longitudinal and
transverse lattice modes, both of which transport charge. After
mixing, one mode is pushed up to $\o_c$, and the other is the
magnetophonon. Tesanovic {\it et al} explicitly computed the
collective modes for the full Hall crystal and showed that there are
only {\it two linearly dispersing gapless modes}. Simply put, for
$n_C=0$, small $q$ oscillations of the lattice produce no charge
motion, and thus no magnetophonon.

In view of the above, it is interesting to ask for the values of the
two integers characterizing the PPHC states. Here there are two spin
flavors, and for charge motion we can treat the two additively. In
this calculation (and in the collective mode calculation that
follows), I employ a trick to keep only the active levels, the $n=1$
majority-spin LL and the $n=0$ minority-spin LL. The trick is to make
$w_c\approx E_Z$ large compared to $e^2/\varepsilon l_0$, which
eliminates LL-mixing. The PPHC state in this regime is adiabatically
connected to the PPHC state at large $r_s$ and $E_Z=0$ (no gaps close
as $E_Z$ is decreased and LL-mixing is introduced). Therefore the two
Chern numbers, and by implication, the structure of the low-energy
collective charge modes, cannot change as one turns on LL-mixing.  The
integers can be easily computed by adapting the results of Tesanovic
{\it et al} for the connection between $n_H$ and
$n_C$\cite{hall-crystal},
\eq
n_C={\phi\over p\phi_0} n_b-{\phi\over\phi_0} n_H
\ee
where $n_b$ is the number of filled nonoverlapping subbands (=4
including both spins for our case), and by borrowing the results of
Yoshioka\cite{yosh} (see also MacDonald\cite{macd2}) for $n_H$ for the
triangular lattice periodic potential. To summarize Yoshioka's
results, if, in a partially filled LL the electrons form an
electron-like Wigner crystal, the contribution to $n_H$ from this LL
is zero, while if they form a hole-like Wigner crytal, they contribute
1 towards $n_H$. Which type of crystal the electrons like to form
depends in turn on the sign of the effective potential $V(\vQ)$. In
our case, it turns out that the $n=0$ minority spin electrons form an
electron-like Wigner crystal, while the $n=1$ majority spin electrons
form a hole-like Wigner crystal (the $n=0$ majority-spin electrons
occupy a full LL, and therefore contribute $n_H=1$). This leads to
$n_H=2$, $n_C=0$, implying that this is a full Hall crystal. Note that
the entire Hall current is carried by majority spin electrons, which
has implications for spin-polarized transport.

Partially polarized square lattice crystalline states also exist. They
are never lower in energy than the triangular ones for the model
interaction I have chosen (though they are quite close). One can
compute the Chern indices for this state as well, by adapting the
results of Hatsugai and Kohmoto\cite{hatsugai}. The difference here is
that a fully gapped half-filled LL with two flux quanta per unit cell
contributes $\pm1$ to $n_H$, depending on the sign of the effective
potential. I find that this state is a {\it partial} Hall crystal,
which has $n_H=1$ and $n_C=2$. This has the amusing feature that its
Hall conductance is $e^2/h$ despite a filling of $\nu=2$!

\begin{figure}
\narrowtext
\epsfxsize=2.4in\epsfysize=2.4in
\hskip 0.3in\epsfbox{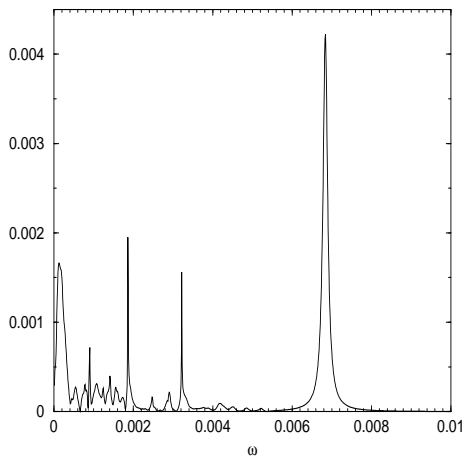}
\vskip 0.15in
\caption{Imaginary part of the charge response function as a 
function of $\o$ for $q=0.03$ in the triangular lattice PPHC. One
optical mode carries most of the spectral weight, and three other
sharp modes are visible at lower energies.
\label{fig2}}
\end{figure}

Let us now turn to collective excitations. In a Wigner crystal one
finds a gapless $q^{3/2}$ magnetophonon, and in a full Hall crystal
one finds two linearly dispersing gapless modes. Thus it is natural to
expect both sets of modes in a partial Hall crystal. Furthermore,
since we have an additional spin degree of freedom there should be
more modes than in the spin-polarized case. I have computed the
collective modes around the HF solution in the time-dependent HF
approximation (TDHF) for both the triangular and square lattice
crystalline states. 

As explained by Cote and MacDonald\cite{cote-macd}, one can reduce the
computation of collective modes to the diagonalization of a large
matrix, from whose eigenvalues and eigenvectors one computes a
response function. The poles of this response function give the
physical collective modes\cite{cote-macd}. The imaginary part of the
charge response function as a function of $\o$ for the triangular
lattice is shown in Figure 2 at $ql_0=0.03$. The feature at $\o\approx
0.007$ is an optical mode, while the sharp features at
$\o\approx0.001,\ 0.002,\ 0.0033$ are gapless linearly dispersing
charge modes. It can be seen that the optical mode has most of the
weight. Now one follows these features as a function of $q$. 
\begin{figure}
\narrowtext
\epsfxsize=2.4in\epsfysize=2.4in
\hskip 0.3in\epsfbox{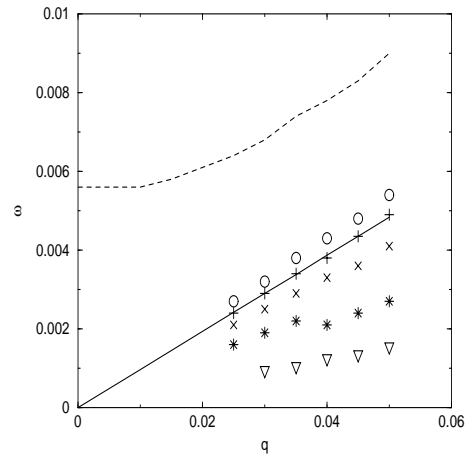}
\vskip 0.15in
\caption{The dispersion of some modes of the triangular lattice PPHC. 
All the linearly dispersing modes extrapolate to $\o=0$ at $q=0$,
showing that they are gapless (see, e.g., the linear fit to the $+$
symbols). Since the spectral weight in the gapless modes decreases as
some power of $q$ for small $q$, their identification above the noise
becomes problematic for $q\le0.02$.
\label{fig3}}
\end{figure}
The resulting set of dispersions for the triangular lattice is shown
in Figure 3. As can be seen, there is no magnetophonon mode dispersing
as $q^{3/2}$, while there are several linearly dispersing collective
modes. All the linearly dispersing modes extrapolate to $\o=0$ at
$q=0$ within error, showing that they are indeed gapless. In Figure 4
we see the corresponding set of dispersions for the square lattice
crystalline state. Here, in addition to the linearly dispersing
gapless modes, the $q^{3/2}$ magnetophonon (symbolized by stars) does
make an appearance, as expected for a partial Hall crystal. 

\begin{figure}
\narrowtext
\epsfxsize=2.4in\epsfysize=2.4in
\hskip 0.3in\epsfbox{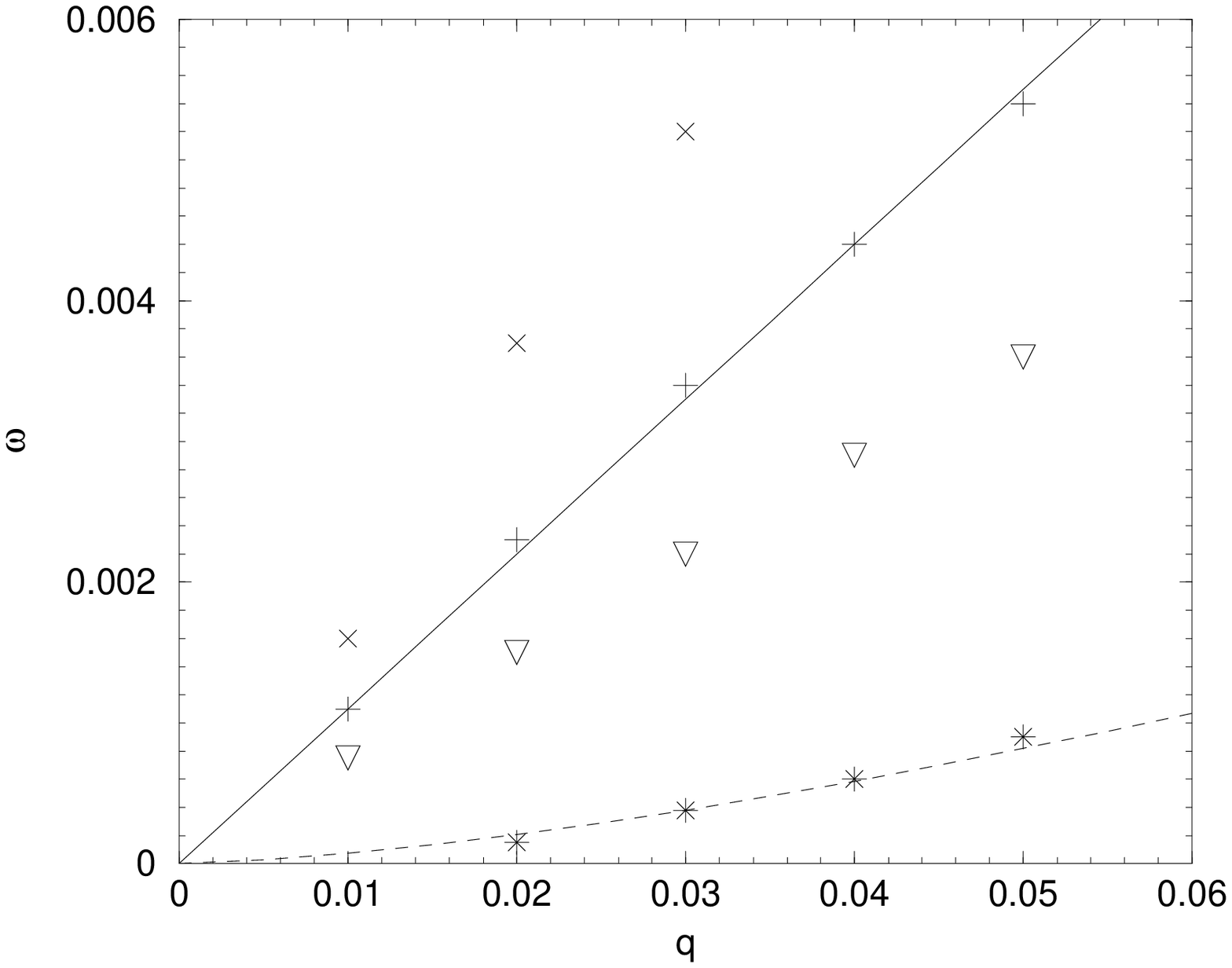}
\vskip 0.15in
\caption{The dispersion of some modes of the square lattice PPHC. 
All the linearly dispersing modes extrapolate to $\o=0$ at $q=0$,
showing that they are gapless (see the linear fit to the $+$
symbols). The mode represented by the star symbol is the magnetophonon
(see the $q^{3/2}$ dashed line fit).
\label{fig4}}
\end{figure}

Let us now turn to other related work. Very recently, Park and
Jain\cite{park-jain} have performed a collective mode analysis (for
zero thickness, $\lambda=0$) of the S and FP states in the $(r_s,E_Z)$
plane. Concentrating on their results for $E_Z=0$, we see that both
states are unstable for $r_s\approx 3$ and greater. To what state
might this instability lead? The PPHC state is definitely not a
candidate at this small $r_s$. I have found another state with equal
occupations of the two spin flavors ($S_{z,total}=0$), with a
triangular lattice density wave with spin-mixing, whose ground state
energy in HF is lower than that of the S and FP states for $1.75\le
r_s \le 2.7$. This state is a full Hall crystal with $n_H=2$ and
$n_C=0$. Strangely enough, this state does not have the full symmetry
of the triangular lattice, implying that the triangular lattice is not
the optimal structure. Some variant of this spin-mixed density wave is
likely to be the ground state for smaller $r_s$. These are likely to
be spin-density waves but total singlets, raising the possibility of
an inhomogeneous quantum Hall {\it antiferromagnet}\cite{dassarma} at
$\nu=2$ (an analogous state at the $\nu=2$ edge has been explored
recently\cite{brey}). The energy of this state is relatively higher
than the PPHC state for larger $r_s$, so I believe that the PPHC state
is still the ground state at large $r_s$. I am intensively exploring
various spin-mixed states at small $r_s$ to resolve this issue.

There are also experimental results on the $\nu=2$ system. Recently,
Eriksson {\it et al}\cite{erik} have measured the collective
excitations of the $\nu=2$ system by inelastic light scattering. They
find that while they see a clear signature of the singlet nature of
the ground state for $r_s\le 3.3$, with a three-fold Zeeman split
spin-density excitation, the situation changes for $r_s\ge 3.3$. Here
they observe  two nondispersing peaks which they interpret as two
roton-like critical points in the dispersion around a singlet state
which has been modified to include Fermi-liquid like
parameters\cite{erik}. They further see the energies of these peaks
decreasing linearly as $r_s$ is increased, suggesting another
transition. It is possible that the first transition is associated
with a transition to the FP or a spin-mixed state, while the second
could be the transition to the PPHC state. Certainly, as one
approaches the transition to the PPHC one expects to see the would-be
linearly dispersing modes soften if the transition is second-order or
weakly first order (see Tesanovic {\it et al}\cite{hall-crystal} for
an example of this). However, further measurements, specifically of
the spin polarization, the Hall conductance, and the collective modes
for $r_s\ge 6$ are needed to uniquely determine the nature of the
state.

In summary, I have shown that there exist partially polarized Hall
crystal states which are likely to be ground states of the $\nu=2$
quantum Hall system at around $r_s\approx 6$. These are HF results,
and subject to the usual caveat: Fluctuations beyond HF can alter the
energies of various states. However, in fully gapped systems such as
these, one expects HF to be not too far off.  The triangular lattice
PPHC state is, to my knowledge, the first realistic full Hall crystal,
and has only linearly dispersing low energy modes for small $q$. The
square lattice crystal state is a partial Hall crystal, with both a
$q^{3/2}$ magnetophonon, and linearly dispersing modes. The square
lattice PPHC state also has an unusual Hall conductance of $e^2/h$
despite a filling of $\nu=2$.

It is a pleasure to thank J.K.Jain, A.H.MacDonald, Z.Tesanovic, and
especially H.A.Fertig for illuminating conversations.

\end{document}